\documentclass{PoS}

\title{Three-dimensional Statistical Jet Fragmentation}

\ShortTitle{3D Statistical Jet Fragmentation}

\author{\speaker{K. Urmossy}$^{\,1,2}$, Z. Xu$^{1,3}$\\
$^{1}$Shandong University, 27 Shanda Nanlu, Licheng, Jinan, P.R. China, 250100\\
$^{2}$on leave from Wigner RCP, 29--33 Konkoly-Thege Miklos Str., Budapest Hungary, H-1121\\
$^{3}$Brookhaven National Laboratory, New York, USA\\
E-mail: \email{karoly.uermoessy@cern.ch}}

\abstract{We reproduce the distribution of the longitudinal and transverse components of momenta of charged hadrons stemming from jets created in proton-proton collisions at $\sqrt s$ = 7 TeV by a statistical fragmentation model. Our hadronisation model is based on microcanonical statistics and negative binomial multiplicity fluctuations. We describe the scale dependence of the fit parameters of the model with formulas obtained by approximating the exact solution of the DGLAP equation in the $\phi^3$ theory with leading order splitting function and 1-loop coupling.}

\FullConference{XXIV International Workshop on Deep-Inelastic Scattering and Related Subjects\\
		11-15 April, 2016\\
		DESY Hamburg, Germany}

\usepackage{graphicx}
\usepackage{amssymb}
\usepackage{epsfig}
\usepackage{amsmath}

\newcommand{\be}{\begin{equation}}
\newcommand{\ee}[1]{\label{#1} \end{equation}}
\newcommand{\ba}{\begin{eqnarray}}
\newcommand{\ea}[1]{\label{#1} \end{eqnarray}}
\newcommand{\nl}{\nonumber \\}

\begin{document}

\section{Introduction}
\label{sec:intro}
As quantum chromo-dynamics (QCD) is not perturbative at small momentum transfere, the process of hadronisation of quarks and gluons created in high-energy collisions is not yet describable directly using first principles. Hadronisation is thus either neglected, or dealt with by means of phenomenological models and empirical functions in the literature. The first way runs under the name 'local parton-hadron duality' (LPHD) conjecture \cite{bib:LPHD} (that is, calculated partonic distributions may be used directly to predict hadron distributions). The other way is to use empirical formulae to describe the fragmentation of each parton species to each hadron species and combine (convolute) them with partonic distributions \cite{bib:AKK}. The distribution of partons in various high-energy processes can be obtained by the resummation of large logarithms \cite{bib:Resum}. For reviews of medium modification of fragmentation functions, see \cite{bib:Arleo} and refs. therein. 

Besides, there are many attempts to obtain fragmentation functions from model calculations. Among all, here, we refere to statistical physical models \cite{bib:Becattini10}-\cite{bib:Begun} which take into account the finiteness of the energy of the created hadronic system, thus, being suitable for the description of momentum fraction and multiplicity distributions of hadrons stemming from jets in $e^+e^-$ and $pp$ collisions. In \cite{bib:UKppFF,bib:UKep}, analitic expressions (cut-power law functions) are obtained for hadron momentum fraction distributions using microcanonical ensemble and superimposed Euler-gamma-type multiplicity fluctuations. Similar distributions are obtained from external volume \cite{bib:Begun}, or temperature \cite{bib:Beck1} fluctuations, or Langevin-equations with multiplicative noise \cite{bib:BiroJako}. Similar distributions are extensively used in high-energy collisions \cite{bib:HIC}.

In Sec.~\ref{sec:relM}, we summarize the statistical hadronisation model introduced in \cite{bib:UKep}. Using this model, in Sec. \ref{sec:exp}, we describe longitudinal and transverse momentum distributions of charged hadrons in jets stemming from proton-proton (pp) collisions at $\sqrt s$ = 7 TeV \cite{bib:atlasFFpp7TeV}. At the end of \ref{sec:exp}, we collect arguments for the representation of hadron distributions in jets of fixed mass-bins instead of fixed momentum-bins to avoid mixing of scales. Besides, we propose an empirical formula for the description of jet mass distributions in pp collisions.

\section{The Model}
\label{sec:relM}
Following \cite{bib:UKep}, we model the created hadrons in a single jet of momentum $P^{\mu}=(E,\mathbf{P})$, mass $M = \sqrt{P_\mu P^\mu}$ and hadron multiplicity $n$ by a microcanonical ensemble. Assuming massless hadrons with momenta $p_i^\mu = (p_i,\mathbf{p}_i)$, the single-particle distribution in the jet is
\be
f_n(p_\mu) \;=\; \frac{(n-1)(n-2)}{\pi M^2} \left(1 - x \right)^{n-3}  \;,
\ee{eq1}
with $x = 2 P_\mu\, p^\mu / M^2$ and normalisation condition $1  \;=\; \int \left(d^3\mathbf{p}/p^0\right) f_n(p_\mu)$. Assuming that hadron multiplicity in jets fluctuates as
\be
\mathcal{P}(n) \;=\; \left(\genfrac{}{}{0pt}{}{n+r-1}{r-1}\right) \tilde{p}^n (1-\tilde{p})^r \;,
\ee{eq2}
the single particle distribution Eq.~(\ref{eq1}) averaged over multiplicity fluctuations becomes
\be
p^0\frac{dN}{d^3\mathbf{p}} \;=\; \sum \mathcal{P}(n) \, n\, f_n(p_\mu) \,= 
 A \left\lbrace \left[1 + \frac{q-1}{\tau}\,x \right]^{-1/(q-1)} - \left[1 + \frac{q-1}{\tau}\right]^{-1/(q-1)} \right\rbrace \;,
\ee{eq3}
with parameters $A = r(r+1)(r+2) [\tilde{p}/(1-\tilde{p})]^3 \;/\; \pi M^2$, $q=1+1/(r+3)$, $\tau = (1-\tilde{p})/[\tilde{p}(r+3)]$. From Eq.~(\ref{eq1}), it follows that particle momenta are within an ellipsoid with centre $\mathbf{P}/2$, longer axis $2a = E$ and smaller axis $2b = M$. The microcanonical nature is manifest in the feature that when particle momenta reach the surface of the ellipsoid ($x = 1$), Eqs.~(\ref{eq1}) and (\ref{eq3}) become zero. 

A possible estimate for the scale dependence of the parameters of Eq.~(\ref{eq3}) can be obtained \cite{bib:UKep} by prescribing that the first three moments of Eq.~(\ref{eq3}), and those of the exact solution of the DGLAP (Dokshitzer-Gribov-Lipatov-Altereli-Parisi) equation in the $\phi^3$ theory with leading-order splitting functions and 1-loop coupling, be equal. The resulting scale dependence is 
\ba
q(t) &=& \frac{(8q_0-12)(t/t_0)^{a_1} - (9q_0-12)(t/t_0)^{-a_2}}{(6q_0-9)(t/t_0)^{a_1} - (6q_0-8)(t/t_0)^{-a_2}}\;, \nl
\tau(t) &=& \frac{\tau_0}{(6q_0-8)(t/t_0)^{-a_2} - (6q_0-9)(t/t_0)^{a_1}} \;,\nl
A(t) &=& [2-q(t)][3-2q(t)]/\tau^2(t) \;
\ea{eq4}
with $t = \ln(Q^2/\Lambda^2)$, $t_0 = \ln(Q_0^2/\Lambda^2)$, $q_0 = q(t_0)$, $\tau_0 = \tau(t_0)$, $A_0 = A(t_0)$.

\section{Results}
\label{sec:exp}
\begin{figure}
\begin{center}
\includegraphics[width=0.45\textwidth, height=0.4\textheight]{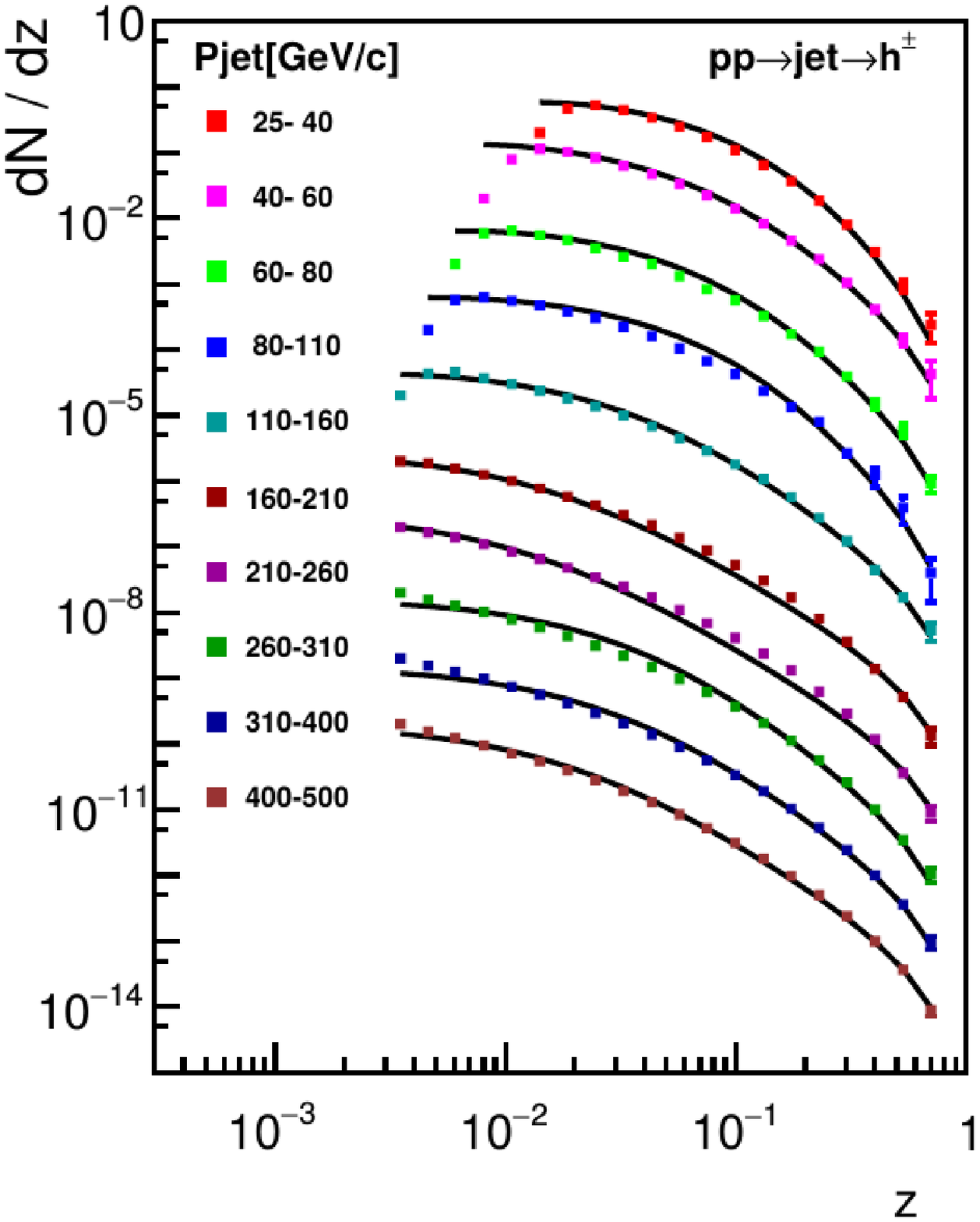} 
\includegraphics[width=0.45\textwidth, height=0.4\textheight]{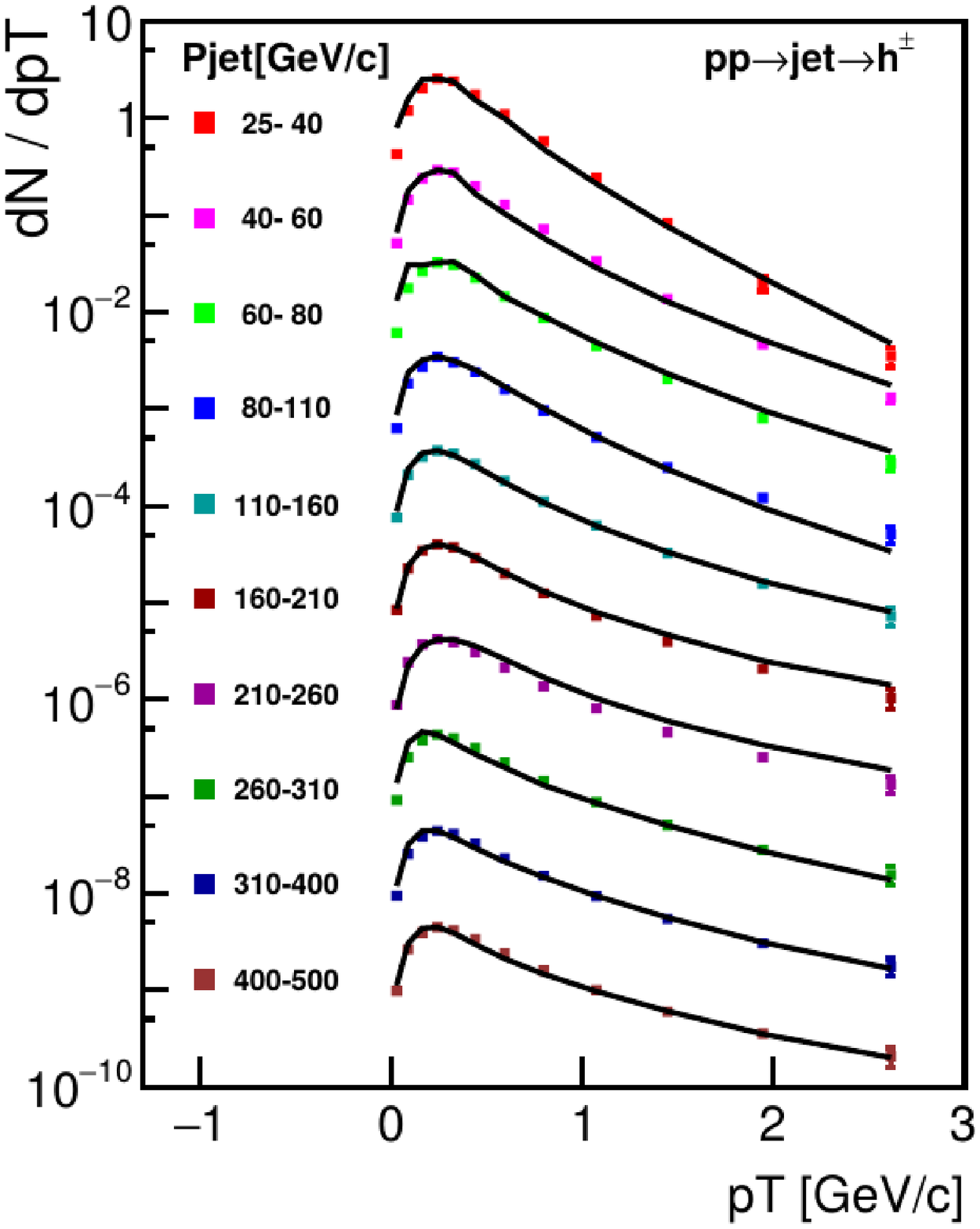} 
\end{center}
\caption{Distributions of the longitudinal (\textbf{left}) and transverse (\textbf{right}) components of momenta of charged hadrons in jets stemming from proton-proton collisions at $\sqrt{s}$ = 7 TeV collision energy. Data are published in \cite{bib:atlasFFpp7TeV}. Curves are fits of Eqs.~(3.1). Data are rescaled for visibility.}
\label{fig:dNdxp}
\end{figure}

In this section, we fit the longitudinal $dN/dz$ ($z=p_z/P$) and transverse $dN/dp_T$ momentum distributions of charged hadrons inside jets stemming from proton-proton collisions at $\sqrt{s}$ = 7 TeV \cite{bib:atlasFFpp7TeV}, using projections of Eq.~(\ref{eq3}):
\ba
\frac{dN}{dp_z} \;&=&\;  2\pi \int\limits_0^{p_T^+} dp_T\, p_T\, \Theta(p_z\tan\vartheta_0 - p_T)\, p^0\frac{dN}{d^3\mathbf{p}} \;,\nl 
\frac{dN}{dp_T} \;&=&\;  2\pi p_T \int\limits_{p_z^-}^{p_z^+} dp_z\, \Theta(p_z\tan\vartheta_0 - p_T)\,\Theta(p_z - p_z^{cut})\, p^0\frac{dN}{d^3\mathbf{p}} \;. 
\ea{eq5}
The limits of the phasespace in this statistical model are $p^{+}_T = \frac{M}{2}\sqrt{1 - \left(\frac{p_z - P/2}{E/2} \right)^2}$ and $p_z^\pm = \frac{P}{2} \pm \sqrt{\left(\frac{E}{2}\right)^2 - \left(\frac{E p_T}{M} \right)^2}$, thus, $p_z \in \left[\frac{P-E}{2}, \frac{P+E}{2} \right]$ and $p_T \in \left[-\frac{M}{2}, \frac{M}{2} \right]$. In the analysis, a jet-cone opening angle $\vartheta_c$ = 0.6 and low momentum cut-off $p_z^{cut}$ = 0.5 GeV/c were used. Though Eq.~(\ref{eq5}) is not in perfect agreement with experimental data (see Fig.~\ref{fig:dNdxp}), this statistical model - despite of its simplicity - catches the main trend of the measured hadron distributions.

\begin{figure}
\begin{center}
\includegraphics[width=0.45\textwidth, height=0.35\textheight]{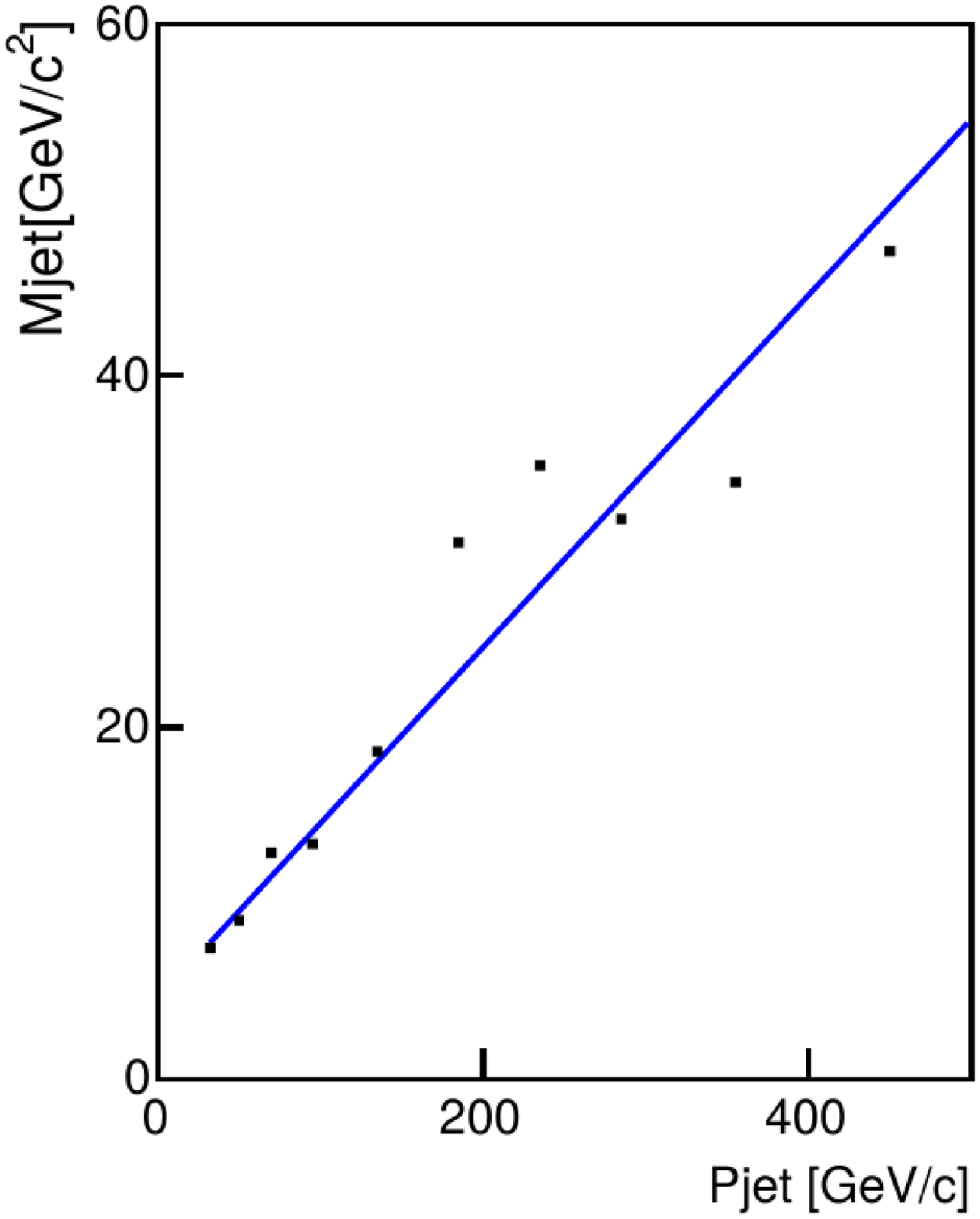} 
\includegraphics[width=0.45\textwidth, height=0.35\textheight]{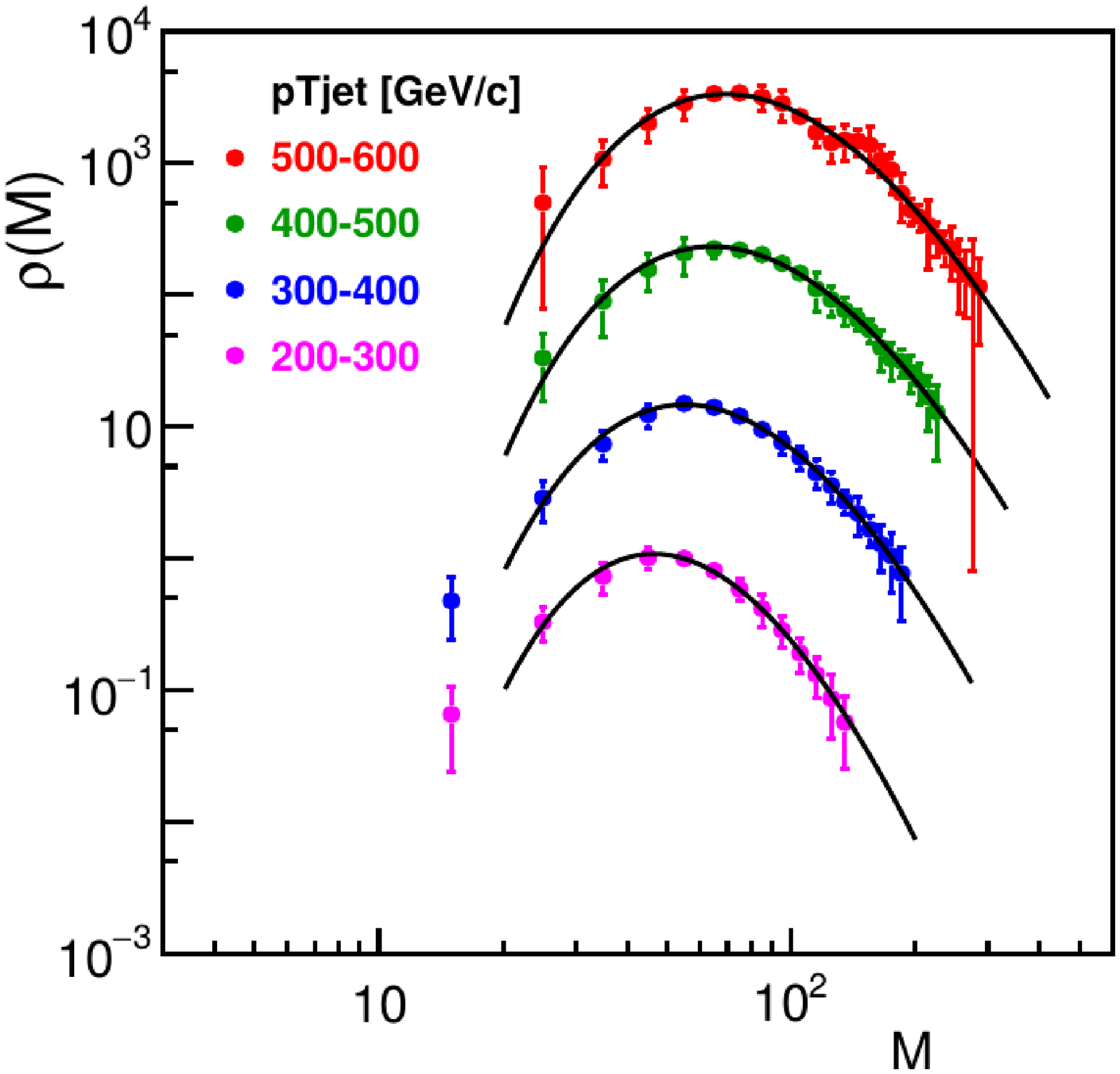} 
\end{center}
\caption{\textbf{Left,} dependence of the fitted characteristic jet mass $M_{jet}$ on the jet 3-momentum $P_{jet}$ (errors are not shown). \textbf{Right,} measured distributions of the mass of jets of various transverse momenta $P_{Tjet}$ \cite{bib:atlasM} along with fits of the empirical formula Eq.~(3.2).\label{fig:M}}
\end{figure}

In the fitting procedure, the $q,\tau$ and $M_{jet}$ parameters were obained for the various $P_{jet}$ datasets. The fitted values of the characteristic jet mass shows a growing tendency as a function of $P_{jet}$ (Fig.~\ref{fig:M}). This can be described by a linear function $M_{jet} = M_0 + E_{jet}/E_0$, with $M_0 = (4.5\pm0.6)$ GeV/$c^2$ and $E_0 = (10.0\pm7.2)$ GeV.

\begin{figure}
\begin{center}
\includegraphics[width=0.45\textwidth, height=0.4\textheight]{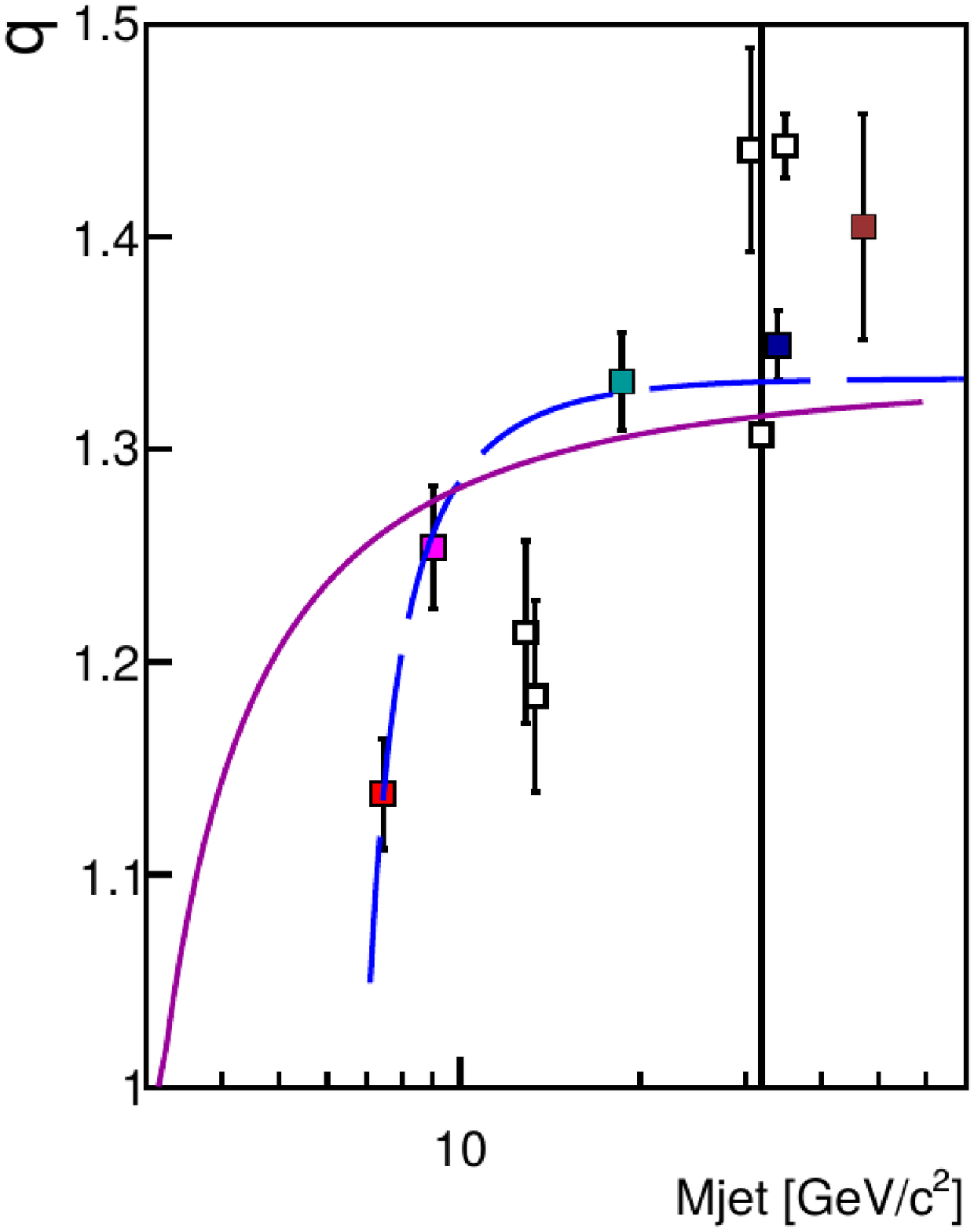} 
\includegraphics[width=0.45\textwidth, height=0.4\textheight]{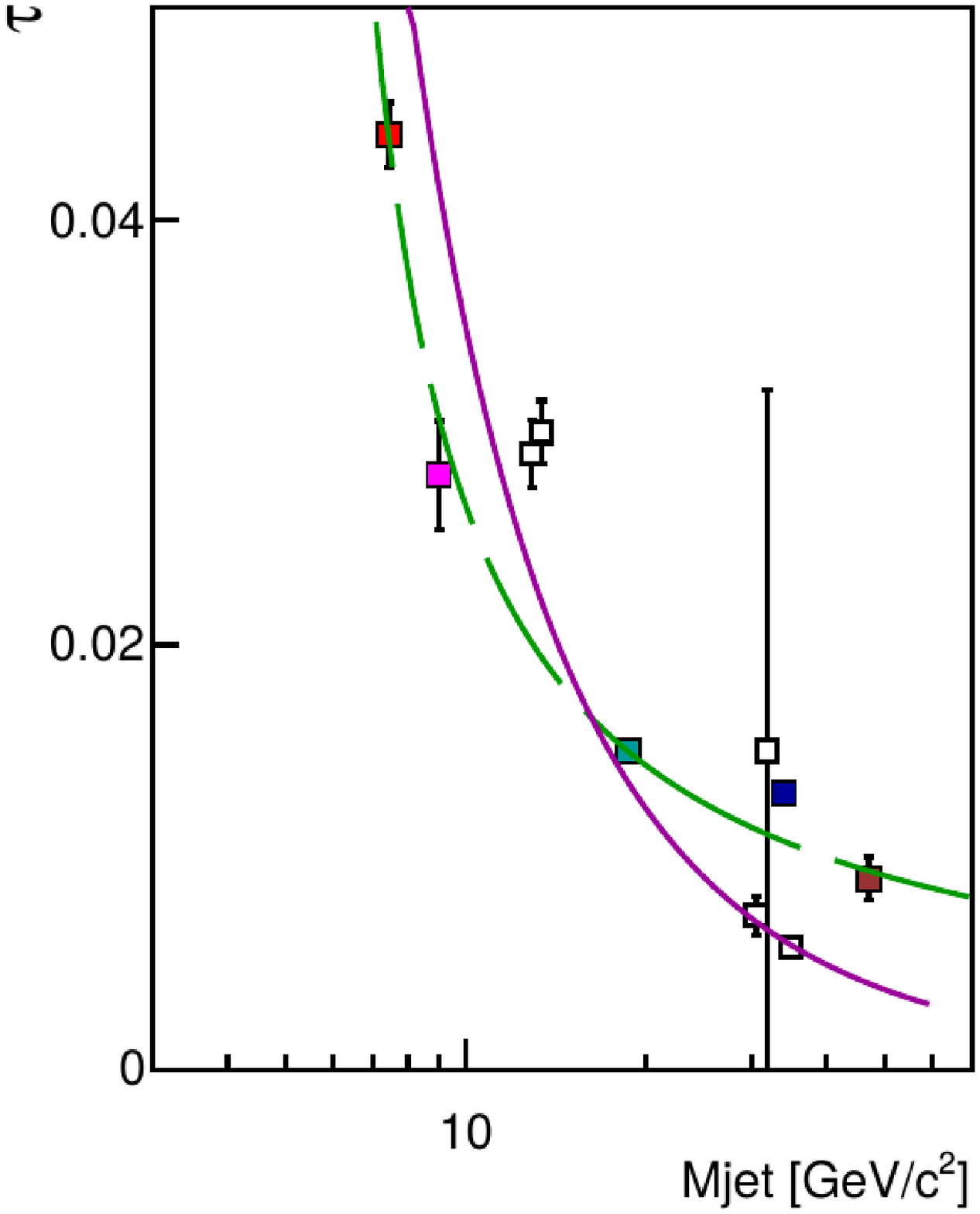} 
\end{center}
\caption{Dependence of the fitted values of the $q$ (\textbf{left}) and $\tau$ (\textbf{right}) parameters of Eq.~(3.1) on the characteristic jet mass $M_{jet}$. Curves are fits of $q\left(M^2_{jet}\right)$ and $\tau\left(M^2_{jet}\right)$ from Eq.~(2.4) with $Q^2 = M_{jet}^2$. See text for more explanations.}
\label{fig:q}
\end{figure}

Dependence of fit parameters $q,\tau$ on $M_{jet}$ are shown in Fig.~\ref{fig:q}. Full symbols correspond to better quality fits. Dashed lines come from fitting the first and second lines of Eq.~(\ref{eq4}) to the $q$ and $\tau$ values (full symbols only) independently, while, solid lines show the result of a simultanous fit. In both cases, the scale was taken to be $Q = M_{jet}$. While the separate fits describe the scale evolution of parameters (full symbols) nicely, the simultanous fit only catches a rough trend. Nevertheless, the $q(M^2_{jet})$ trend suggests that the lighter the jet, the smaller the $q$. Note, that as $q\rightarrow 1$, the multiplicity distribution Eq.~(\ref{eq2}) tends to the Poissonean $\mathcal{P}(n)\rightarrow (1/\tau)^n\exp\lbrace-1/\tau\rbrace /n!$, while, the hadron distribution  Eq.~(\ref{eq3}) tends to the Boltzmann-Gibbs $p^0 \left(dN/d^3\mathbf{p}\right) \rightarrow A\exp\lbrace-x/\tau \rbrace$.

\subsection*{Jet Mass Fluctuations}

It is important to note that the mass of jets in pp collisions has a broad distribution even inside a certain $P_{jet}$ bin (Fig.~\ref{fig:M}, right panel). This distribution can be described by the empirical function
\be
\rho(M) \;=\;  \frac{\ln^b(M/\mu_0)}{M^c} \;.
\ee{eq6}
If we accept that the proper scale of the fragmentation function is the jet mass $Q^2 = P_\mu P^\mu$, then the above analysed hadron distributions in each $P_{jet}$ bin originate from jets of various $Q$ scales (masses). Naturally comes the question, would it not be better to present hadron distributions in jets of fixed mass-bins instead of in fixed $P_{jet}$-bins? This way, the mixing of different $Q$ (mass) scales would be avoided.

In this work, we have only fitted an average, characteristic value for the jet mass. In a future work we will also take jet mass fluctuations into account and might get better description of measured data.


\section*{Acknowledgement}
These research results were sponsored by the China/Shandong University International Postdoctoral Exchange Program. This paper was also supported by the Hungarian OTKA Grant K104260. Author K. U. is thankful to David d'Enterria for discussions.


\end{document}